\documentstyle[prl,aps]{revtex}

\begin{document}

\draft

\twocolumn[\hsize\textwidth\columnwidth\hsize\csname
@twocolumnfalse\endcsname
 
\title{ Intermediate Valence Model for Tl$_{2}$Mn$_{2}$O$_{7}$ }

\author {C.I. Ventura and B.R. Alascio~$^*$}

\address {Centro At\'omico Bariloche, 8400-Bariloche, Argentina.}

\maketitle

\begin{abstract}
{\small
There have been speculations about the need to find a new mechanism to
explain the colossal magnetoresistance exhibited by this material, 
having pyrochlore
structure and thus differing  structurally and electronically 
from the manganites. We will report here our transport results based on a
two band model, with conduction electrons and intermediate
valence ions fluctuating between two magnetic configurations. The
model has been previously employed to understand transport and 
thermodynamical properties of intermediate valence Tm compounds and, 
in its periodic version, to analize the phase diagram. The results
obtained with this model for the transport properties   
of Tl$_{2}$Mn$_{2}$O$_{7}$ are in good qualitative agreement 
with the experimental results. }
 
\end{abstract}

\vskip2pc] \narrowtext




	The strong correlation between transport properties and magnetism
in perovskite manganese oxides like La$_{1-x}$M$_{x}$MnO$_{3}$
(M=Ca,Sr,Ba) has been known for many years \cite{1}. Shortly after the
discovery of these materials, the double exchange mechanism was
proposed \cite{2} to describe the interactions between the Mn ions
which, due to the divalent substitution for La, are in a mixed valence
state. Electronic carrier hopping between heterovalent Mn pairs 
(Mn$^{3+}$-Mn$^{4+}$) is enhanced by the mutual alignment of the two
magnetic moments. Thus the resistivity will depend on the spin
disorder and is expected to display pronounced features at the
ferromagnetic ordering transition temperature (T$_{c}$). On
application of a magnetic field, which tends to align the local spins,
the resistivity is expected to decrease.

	Recently, colossal
magnetoresistance (denoted MR hereafter) was observed near the 
ferromagnetic ordering
temperature \cite{3} and the interest in the study of these materials
was renewed. At present, there is disagreement on whether theoretical
models based only on the double exchange mechanism can account
quantitatively  for the observed transport and magnetic properties of 
the manganese perovskites. It has been proposed that other ingredients
such as disorder \cite{4}, Jahn Teller distorsions \cite{4'}, etc.
should be included and would play an important role.

	In 1996 colossal magnetoresistance was observed for the
non-perovskite Tl$_{2}$Mn$_{2}$O$_{7}$ compound \cite{5,6,7}. This
material has the pyrochlore A$_{2}$B$_{2}$O$_{7}$ structure \cite{7'},
consisting of AO$_{8}$ cubes and BO$_{6}$ octahedra linked to form a
three-dimensional network of corner-sharing tetrahedra. This 
tetrahedral MnO$_{6}$ (B=Mn) network differentiates the pyrochlores
from the perovskite ABO$_{3}$ structure, with a cubic MnO$_{6}$
network.The pyrochlore structure is face-centered cubic with 8 formula units
per unit cell.
Tl$_{2}$Mn$_{2}$O$_{7}$ undergoes a ferromagnetic transition
 with T$_{c}$$\sim$ 140 K \cite{5,6,7}. Below T$_{c}$ the compound is
ferromagnetic, whereas above T$_{c}$ it is paramagnetic. The 
magnetoresistance maximum around the ferromagnetic ordering
temperature is similar to that obtained for the manganese perovskites.
Mostly due to the absence of evidence for significant doping in the
pyrochlore Mn-O sublattice (to produce the mixed valence responsible
for double exchange in perovskite manganese oxides about 20-45$\%$
doping is needed), and due to the tendency of Tl to form 6-s
conduction bands (unlike the perovskites where the rare-earth levels
are electronically inactive), it has been put to question whether a
double exchange mechanism similar to that of perovskites can account
for the experimental results \cite{5,6,7}. Hall data \cite{5} show
a very small number of n-type carriers ( $\sim$ 0.005 conduction 
electrons per formula unit) and this would seem to indicate
a very small doping into the Mn$^{4+}$ state \cite{5,7}. 
Among possible explanations, the authors of
Refs.[6,8] mention that such Hall data could result from a small number
of carriers in the Tl 6-s band, so that assuming
Tl$_{2-x}^{3+}$Tl$_{x}^{2+}$Mn$_{2-x}^{4+}$Mn$_{x}^{5+}$O$_{7}$ with 
x$\sim$ 0.005 the data could be accounted for.

	Based on these facts we decided to explore the suitability of 
the intermediate valence (IV) model now to be introduced for 
Tl$_{2}$Mn$_{2}$O$_{7}$. The model was proposed originally for the
study of Tm compounds \cite{8,9}. The exactly solvable impurity model
for valence fluctuations between two magnetic configurations was shown
\cite{8} to describe most of the peculiar features of the magnetic
properties of paramagnetic intermediate valence Tm compounds.  The
impurity model resistivity exhibits an explicit quadratic dependence
with the magnetization\cite{8}. Such a behaviour has also been found in
transport experiments slightly above T$_{c}$ for colossal MR
pyrochlores \cite{5} and manganese perovskites \cite{10}. With
the periodic array of IV ions version of the model \cite{9} the effect
of the interactions between Tm ions was studied, obtaining the T=O
phase diagram, specific heat and magnetic susceptibility for the
paramagnetic phase in coherent potential approximation (CPA) and the 
neutron scattering spectrum. For manganese perovskites a similar model
has been considered \cite{11} to propose the possibility of a
metal-insulator transition.

	We will now present the model employed to describe a
periodic lattice of intermediate valence (IV) ions, which fluctuate
between two magnetic configurations associated to single (S=1/2) or
double occupation (S=1) of the ion, hybridized to a band of conduction
states. The hamiltonian considered is \cite{9}:
\begin{eqnarray}
H \, & \, = \, & \, H_{L} \, + \, H_{c} \, + \, H_{H} \,
\end{eqnarray}
where:
\begin{eqnarray}
H_{L} \, & \, = \, & \, \sum_{j} \left( E_{\uparrow}   \mid j
\uparrow > 
<  j \uparrow \mid   
+  E_{\downarrow}   \mid j \downarrow
> < j \downarrow \mid  \right) \nonumber \\ 
\, &  & \, + \left( E_{+}   \mid j + > 
<  j + \mid  +  E_{-}   \mid j - > < j - \mid \right) \, , 
\nonumber \\ 
H_{c} \, & \, = \, & \,\sum_{k, \sigma} \epsilon_{k,\sigma} \, 
c^{+}_{k, \sigma} c_{k, \sigma}  \, , \nonumber \\
H_{H}  & \, = \, & \sum_{i,j}  V_{i,j} \left(  \mid j + > < j
\uparrow \mid 
c_{i,\uparrow}  +   \mid j - > < j \downarrow \mid 
c_{i,\downarrow} \right) + H.c.  \nonumber 
\end{eqnarray}

The IV ions, which for Tl$_{2}$Mn$_{2}$O$_{7}$ we would identify with 
the Mn ions,
are represented by H$_{L}$, which describes the S=1/2
magnetic configuration at site j through states $\mid j \sigma >$ ( $\sigma$ =
$\uparrow, \downarrow$ ), with energies E$_{\sigma}$ split in the
presence of a magnetic field B according to:
\begin{eqnarray}
E_{\uparrow (\downarrow)} \,   \, = \,   \, E - (+) \mu_{0} B . 
\label{eupdown}
\end{eqnarray}
The S=1 magnetic configuration is considered in the highly anisotropic
limit, where the S$_{z}$ = 0 state is projected out of the subspace of
interest as in Refs.[10,11]. The S=1 states at site j are
represented by $\mid j s >$ (s=+,$-$) and energies E$_{s}$, split by the
magnetic field as:
\begin{eqnarray}
E_{ \pm } \,   \, = \,   \, E + \Delta \mp \mu_{1} B .
\label{emm}
\end{eqnarray}

H$_{c}$ describes the conduction band, which for
Tl$_{2}$Mn$_{2}$O$_{7}$ we would identify with the Tl 6s conduction
band. Through hybridization 
with the conduction band, H$_{H}$ describes valence
fluctuations between the two magnetic connfigurations at one site.
For example, through promotion of a spin up electron into the
conduction band at site j the IV ion passes from state $\mid j +>$ to
state $\mid j \uparrow>$. Notice that the highly anisotropic limit
considered inhibits any spin flip scattering of conduction electrons,
so that the direction of the local spin at each site is conserved.
IV ions only hybridize with conduction electrons of parallel spin.

The hamiltonian can be rewritten in terms of the following creation
(and related annihilation) operators for the local orbitals (Mn 3-d 
orbitals, in this case) \cite{9}:
\begin{eqnarray}
d^{+}_{j,\uparrow} \, = \, \mid j + > < j \uparrow \mid \; , \nonumber
\\
d^{+}_{j,\downarrow} \, = \, \mid j - > < j \downarrow \mid ,
\end{eqnarray}
for which one has:
\begin{eqnarray}
\left[ d_{i, \uparrow}, d^{+}_{j, \uparrow} \right]_{+} \, = \, 
\delta_{i,j} \left( P_{i,\uparrow} + P_{i,+} \right) \; , 
\nonumber \\
\left[ d_{i, \downarrow}, d^{+}_{j, \downarrow} \right]_{+} \, = \, 
\delta_{i,j} \left( P_{i,\downarrow} + P_{i,-} \right) \, , \nonumber \\
P_{i,+} + P_{i,\uparrow} + P_{i,-} + P_{i,\downarrow} \, = \, 1. \nonumber
\end{eqnarray}
where: P$_{j,\alpha}$ = $ \mid j \alpha > < j \alpha \mid $ are
projection operators onto the local magnetic configuration states. 
The local hamiltonian now reads:
\begin{eqnarray}
H_{L} \, & \, = \, & \,  \left( \Delta - \mu_{D} B \right) \sum_{j} 
d^{+}_{j,\uparrow} d_{j,\uparrow} + \left( \Delta + \mu_{D} B \right)
\sum_{j} d^{+}_{j,\downarrow} d_{j,\downarrow} , \nonumber \\
\mu_{D} \, & \,  = \, & \, \mu_{1} - \mu_{0}. 
\end{eqnarray}

	Due to the type of hybridization present, one can now consider the
problem as described by two separate parts \cite{9}. Given a certain
configuration for the occupation of the local orbitals at all sites 
with spin up or down electrons, the spin up conduction electrons will
only hybridize with those IV ions occupied by spin up electrons (i.e. in $
\uparrow$ or $+$ local states). One could simulate this by including a
very high local correlation energy (U $ \rightarrow \infty$) to be
paid in the event of mixing with ions occupied by opposite spin
electrons. Concretely, we can take \cite{9}:
\begin{eqnarray}
H \,& \,  = \, & \, H_{\uparrow} + H_{\downarrow} , \\
H_{\uparrow} \, & \,  = \, & \, \left( \Delta - \mu_{D} B \right)
\sum_{j 
\in \uparrow} 
d^{+}_{j,\uparrow} d_{j,\uparrow} \nonumber \\ \, & \, & \, 
 + U \sum_{j \in \downarrow} 
d^{+}_{j,\uparrow} d_{j,\uparrow} + \sum_{k} \epsilon _{k} 
c^{+}_{k,\uparrow} c_{k,\uparrow} \nonumber \\
\, & \, & \, + \sum_{i,j} \left( V_{i,j} d^{+}_{j,\uparrow} c_{i,\uparrow} +
H.c. \right). \nonumber
\end{eqnarray}
H$_{\downarrow}$ is analogous to H$_{\uparrow}$, one having only 
to reverse the sign of B and spin directions. We will ignore the
split of the conduction band energies in the presence of the magnetic
field B, and take for the hybridization: V$_{i,j}$ = V $\delta_{i,j}$.

	Given a certain configuration of up and down spin occupations
of the sites, one can now solve two separate alloy problems, described
by H$_{\uparrow}$ and H$_{\downarrow}$ respectively. These we solved in
CPA approximation as in Ref.\cite{9}, introducing an effective
diagonal self-energy for the local orbitals, $\Sigma_{(d) \sigma}(\omega)$
for the H$_{\sigma}$ alloy problem, through which on average
translational symmetry is restored.
The CPA equation obtained to determine
self-consistently the self-energy
relates it directly to the local Green function for the IV ions:
\begin{eqnarray}
\Sigma_{(d) \uparrow}(\omega) \, = \, \frac{p - 1}{<< d_{j,\uparrow}, 
d^{+}_{j,\uparrow} >>(\omega) } \, , \nonumber \\
\Sigma_{(d) \downarrow}(\omega) \, = \, \frac{- p}{<< d_{j,\downarrow}, 
d^{+}_{j,\downarrow} >>(\omega) } .
\end{eqnarray}
Here p denotes the concentration of spin up sites, and for the densities
of states hold:
\begin{eqnarray}
\int_{-\infty}^{\infty} d\omega  \rho_{c,\sigma}(\omega) \, & \, = \,
1 , \nonumber
\\ \int_{-\infty}^{\infty} d\omega  \rho_{d,\uparrow}(\omega) \, & \,
= \, & \, p \, ,  \;
\int_{-\infty}^{\infty} d\omega  \rho_{d,\downarrow}(\omega) \,  \, =
\,  \, 1 - p.
\end{eqnarray} 

The CPA equation is solved self-consistently with the total number of
particles equation, through which the chemical potential is determined.
    
	To take into account the effect of temperature 
on the magnetization and describe
qualitatively the experimental data in pyrochlores 
\cite{5,7}, we will use a simple Weiss molecular field approximation 
to obtain the magnetization at each temperature \cite{Coqblin}  
and through this the 
concentration p of ions occupied by spin up electrons.   

	Considering now the determination of transport properties, 
using the Kubo formula it has been proved before
\cite{velicky,brouers} 
that no vertex corrections to the conductivity are 
obtained in a model such as this.
Furthermore, in the absence of a direct hopping term between local
orbitals only the conduction band will contribute to the conductivity 
\cite{brouers,sakai}. As in Refs. \cite{velicky,brouers,sakai} one can
obtain the conductivity through Boltzmann equation in the relaxation
time approximation limit as:
\begin{eqnarray}
\sigma \, & \,  = \, & \, \sigma_{c,\uparrow} + \sigma_{c,\downarrow}
, \nonumber
\\  \sigma_{c,\uparrow} \, & \, = \, & \, n_{c} e^{2} \int 
d\omega \left( - \frac{
\partial f (\omega) }{\partial \omega} \right)
\tau^{k}_{c,\uparrow}(\omega) 
\Phi(\omega),
\end{eqnarray}
where f is the Fermi distribution, n$_{c}$ the total number of
carriers per unit volume, and the relaxation time for 
spin up conduction electrons is:
\begin{eqnarray}
\tau^{k}_{c,\uparrow}(\omega) \, = \, \frac{\hbar}{2 \mid Im \Sigma^{k}_{c
\uparrow} (\omega)}
\end{eqnarray}
and: $ \Phi \, = \, \frac{1}{N} \sum_{k} v^{2}_{c}(\epsilon_{k})
\delta(\omega - \epsilon_{k}) $, where $v_{c}(\epsilon_{k})$ is the
the conduction electron velocity.
The extension of these formulas for spin down is straightforward. 

Considering temperatures much lower than the Fermi temperature, one
can approximate: $\Phi (\omega) \sim v_{F}^{2} \rho_{c}^{(0)}(\omega)$,
 being $ v_{F} $ the Fermi velocity. The 
results presented here were obtained assuming
for simplicity a semielliptic bare density of states for the
conduction band $\rho_{c}^{0}(\omega)$, in which case 
the k-dependent self-energy for
conduction electrons is related to the effective medium CPA
self-energy through: 
\begin{eqnarray}
\Sigma^{k}_{c \uparrow}(\omega) \, & \,  = \, & \, \frac{V^{2}}{\omega
- \sigma_{(d)\uparrow}(\omega)} \,  , \nonumber \\ 
\sigma_{(d)\uparrow}(\omega) \, & \, = \, & \, \Delta - \mu_{D} B + 
\Sigma_{(d)\uparrow} (\omega).
\end{eqnarray}

	In Fig.\,\ref{fig:resred} we show the results obtained for 
the resistivity 
employing this model with parameters: W=6eV for the semielliptic bare 
conduction half-bandwidth(centering the conduction band at the
origin), $E=0$, $\Delta$ = -4.8 eV, V = 0.6eV and total
number of particles per site n = 1.075. To reproduce qualitatively
the experimental magnetization data \cite{5,7} we take T$_{c}$ as 142K 
and a saturation value 
of 3$\mu_{B}$ (like for free Mn$^{4+}$ ions) in the Weiss
approximation, as well as $\mu_{D}= 1 \mu_{B}$.  
In Fig.\,\ref{fig:dered} we show the gap in the 
CPA spin up densities of states obtained at T=0 with those 
parameters. 
At temperatures above Tc there still are gaps present (or at least 
pseudogaps, as is the case for the spin down bands at the higher 
magnetic fields shown: B=4T,8T).
A filling of n=1.075 corresponds to the Fermi level placed slightly
above the bottom of the upper bands ($\mu \sim -4.51 eV$). 
The resistivity curves in 
Fig.\,\ref{fig:resred} have a behaviour around T$_{c}$ similar to that
found in experiments \cite{5,6,7}. An order of magnitude estimate for 
our resistivity results for n= 1.075 is compatible with the values 
experimentally found (see Fig.\,\ref{fig:resred}), in any case the 
absolute value of resistivity we obtain depends on the filling.
 In Fig.\,\ref{fig:mrred} we plot
magnetoresistance results obtained with the same parameters. Here 
our results exhibit a difference in value between the MR maxima around
T$_{c}$ for 
different magnetic fields and  
a crossing at higher temperatures   
of the MR curves obtained for those B, such as are present  
in the experimental data \cite{7}. 
The description of the main features exhibited by
transport measurements around T$_{c}$ is quite remarkable, considering
the simplifications adopted here. Moreover, with the parameters
employed (and due to the resulting presence of gaps or pseudogaps
slightly below the Fermi level) we have a small 
effective number of electrons participating in the transport (mostly
the few Tl conduction electrons above the gap) 
like Hall experiments indicate \cite{5,7}.
Nevertheless it is the total number of conduction electrons (which includes
those below the gap and with our parameters is about 0.15, ten times
the number of carriers above the gap)
which would represent the real doping (x) into the Mn$^{4+}$ state 
to take into consideration. This could solve the difficulties with   
Hall data which are mentioned in Refs.\cite{5,7}. 
 
	We will now briefly comment on the transport
results obtained with the impurity version of the model discussed
above \cite{8}. Using parameters for the impurity in accordance with
those employed for the lattice, and a filling such that the Fermi level falls
slightly above the peak of the impurity density of states, one
can obtain transport results which are very similar to those 
presented above. Nevertheless, it would be hard to reconcile the impurity
picture with the Hall data indicating a very small number of carriers 
effective in transport \cite{5,7}.

To conclude we will indicate that the agreement between our results
and the experimental data available on colossal MR pyrochlore 
Tl$_{2}$Mn$_{2}$O$_{7}$ is indeed very reasonable, and we believe that
it can certainly be improved by fine tuning of the parameters of the
model. The presence of gaps or pseudogaps in the electronic structure,
such as those considered here in the periodic model, not only could
solve the problems posed by the Hall data which are mentioned by the
authors of Refs.\cite{5,7}. But their presence should cause observable
effects in other experiments, such as spin polarized tunneling and optical
properties which would be interesting to investigate.

\vspace{1cm}

{\bf Figure Captions:}
\begin{figure}
\caption{Periodic IV model. Resistivity as a function of 
temperature for magnetic fields: B= 0,1,4 and 8T. Parameters: W= 6 eV, 
E=0, $\Delta=$ -4.8 eV, V= 0.6 eV, n= 1.075, Tc= 142 K, M$_{sat}$=
3$\mu_{B}$, $\mu_{D}$= 1 $\mu_{B}$. $c = (eV)^{2}/ [n_{c} e^{2}
v_{F}^2 \hbar]$, e.g. $\, c \sim 0.1 \, \Omega cm$ for $n_{c} \sim
10^{21}/cm^{3}$ and $v_{F} \sim 10^{7}cm/s$.}
\label{fig:resred}
\end{figure} 
\begin{figure}
\caption{Periodic IV model. Densities of states as a
function of energy at T=0, B=1T, parameters as in Fig.(1).}
\label{fig:dered}
\end{figure}
\begin{figure}
\caption{Periodic IV model. Magnetoresistance as a function
of temperature, from curves of Fig.(1). 
Full line: $\frac{\rho_{1T}(T)
-\rho_{4T}(T)}{\rho_{4T}(T)}$; dashed line: $\frac{\rho_{4T}(T)
-\rho_{8T}(T)}{\rho_{4T}(T)}$.}
\label{fig:mrred}
\end{figure}

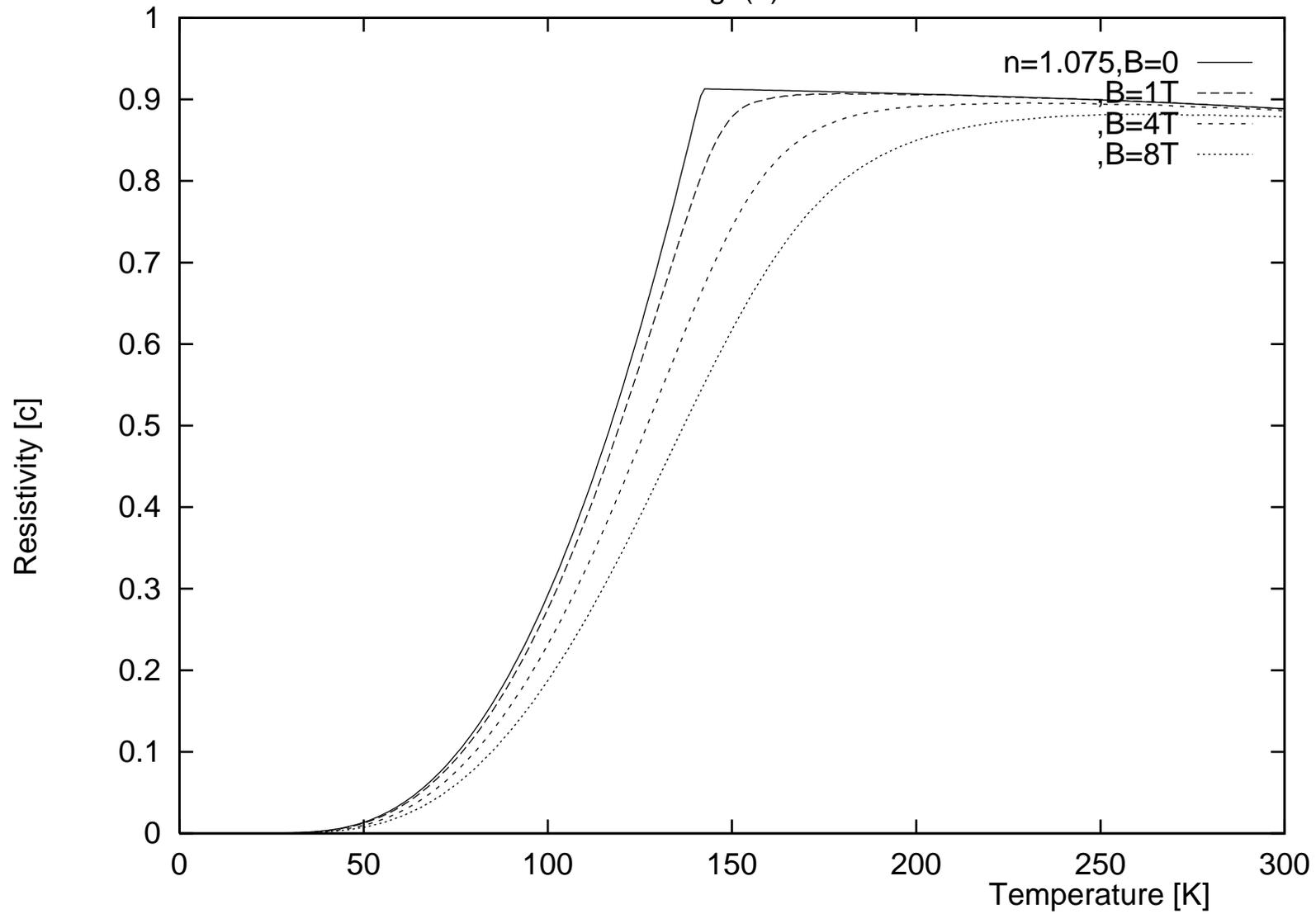
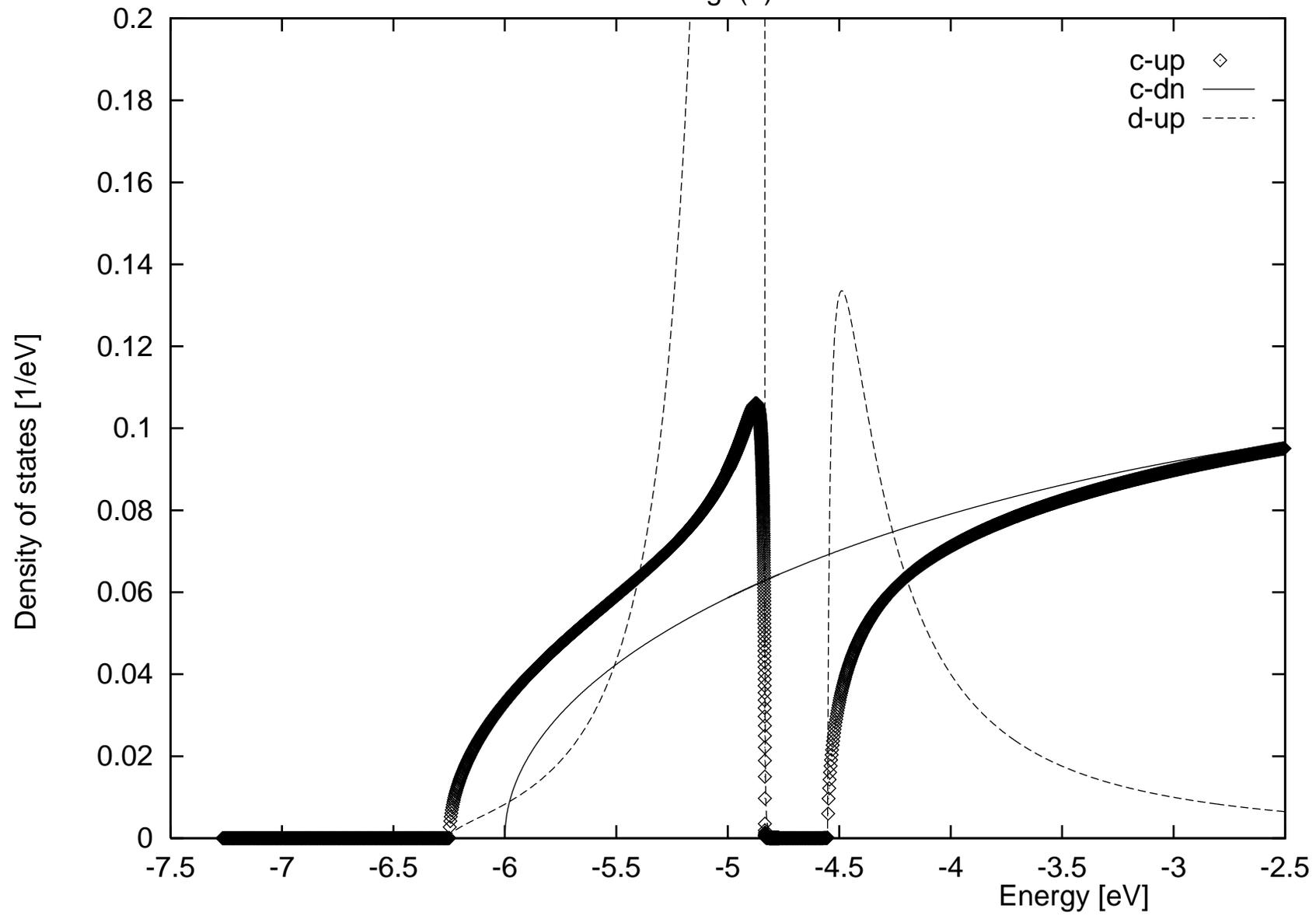
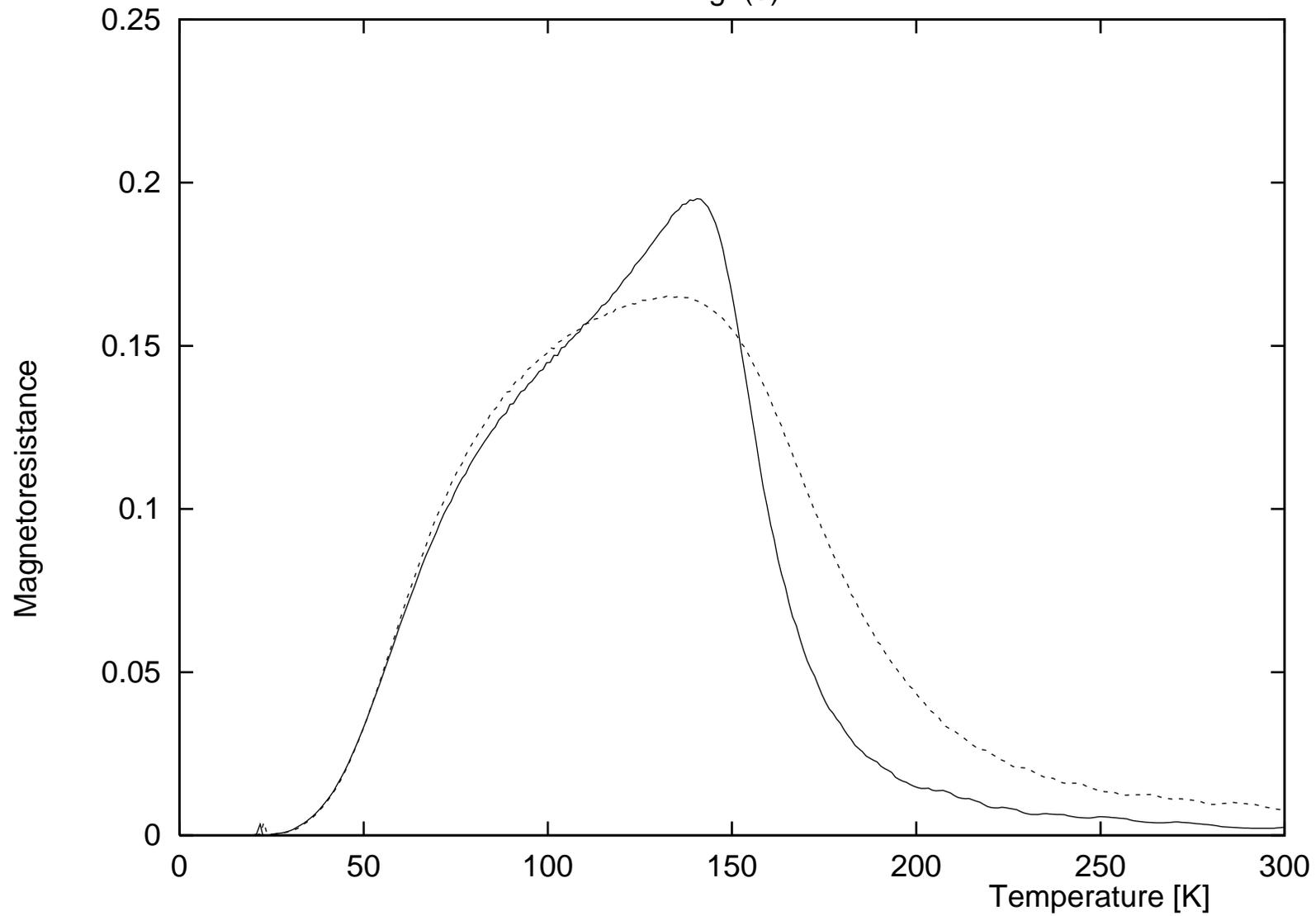
\end{document}